\documentstyle[12pt]{article}
\def\bbox{{\,\lower0.9pt\vbox{\hrule \hbox{\vrule height 0.2 cm

\hskip 0.2 cm

\vrule  height 0.2 cm}\hrule}\,}}

\def\bbox{{\,\lower0.9pt\vbox{\hrule \hbox{\vrule height 0.2 cm

\hskip 0.2 cm

\vrule  height 0.2 cm}\hrule}\,}}

\begin{document}
\setlength{\unitlength}{1mm}
\title{{\hfill {\small } } \vspace*{2cm} \\
Energy, Hamiltonian, Noether Charge, \\
and Black Holes}
\author{\\
Dmitri V. Fursaev\thanks{e-mail:
fursaev@thsun1.jinr.ru}
\date{}}
\maketitle
\noindent  {
{\em Joint Institute for
Nuclear Research,
Bogoliubov Laboratory of Theoretical Physics, \\
141 980 Dubna, Russia}
}
\bigskip

\begin{abstract}
It is shown that in general
the energy ${\cal E}$ and the Hamiltonian
${\cal H}$ of
matter fields on the black hole exterior
play different roles. ${\cal H}$ is a
generator of the time evolution along the Killing time while ${\cal E}$
enters the first law of black hole thermodynamics.
For non-minimally coupled fields the difference
${\cal H}-{\cal E}$ is not zero and is a Noether charge $Q$
analogous to that
introduced by Wald to define the black hole entropy.
If fields vanish
at the spatial boundary, $Q$ is
reduced to an integral over the horizon.  The analysis is
carried out and an explicit expression for $Q$ is found for general
diffeomorphism invariant theories.
As an extension of the results by Wald et al,
the first law of black hole thermodynamics
is derived for arbitrary weak
matter fields.
\end{abstract}

\bigskip\bigskip

\baselineskip=.6cm

\newpage

\section{Introduction}
\setcounter{equation}0

There may be two definitions of the energy of
matter fields on an external time independent background.
The first one defines the energy in terms
of the stress-energy tensor $T_{\mu\nu}$ as
\begin{equation}\label{i1}
{\cal E}_{(m)}=-\int_{\Sigma_t} \sqrt{-g} d^3x T^0_0~~~,
\end{equation}
where $\Sigma_t$ is a hypersurface of a constant time $t=const$,
$g$ is the determinant of the metric tensor $g_{\mu\nu}$ of
the background space-time\footnote{We follow conventions of book
\cite{MTW}}. The definition of the stress-energy tensor is standard
\begin{equation}\label{i2}
T^{\mu\nu}={2 \over \sqrt{-g}}
{\delta I_{(m)} \over \delta g_{\mu\nu}}~~~,
\end{equation}
in terms of the action $I_{(m)}$ of matter fields. Another
possibility is to identify the energy with the Hamiltonian
\begin{equation}\label{i3}
{\cal H}_{(m)}=\int_{\Sigma_t}
\sqrt{-g} d^3x \left[ {\partial L_{(m)} \over \partial \dot{\phi}}
\dot{\phi}-L_{(m)}\right]~~~,
\end{equation}
where $L_{(m)}$ is the Lagrangian of
fields $\phi$, and $\dot{\phi}$
are the time derivatives of $\phi$.

The energies ${\cal E}_{(m)}$ and ${\cal H}_{(m)}$ do not depend on $t$
(or on the choice of the hypersurface $\Sigma_t$)
provided $\phi$ obey the equations of motion and vanish
at the spatial boundary $C_t$ of $\Sigma_t$ or at asymptotic
infinity.
It can be shown that ${\cal E}_{(m)}$ and ${\cal H}_{(m)}$ either coincide or
differ by a surface term
on $C_t$.
In the most physical situations, however, the boundary
conditions can be chosen in such a way to eliminate the
boundary terms and make ${\cal E}_{(m)}$ and ${\cal H}_{(m)}$ equal.

The black holes are an exclusion from this rule.
For a system in the black hole
exterior, the surfaces $\Sigma_t$ meet at the bifurcation surface
$\Sigma$ of the black hole horizon. It is important
that $\Sigma$ is an internal boundary of $\Sigma_t$ where
fields obey no conditions but
regularity. Thus, if ${\cal E}_{(m)}$ and ${\cal H}_{(m)}$ differ by a surface
term the contribution from $\Sigma$ cannot be eliminated.
On a static black hole background one can write
\begin{equation}\label{i4}
{\cal H}_{(m)}-{\cal E}_{(m)}={\kappa \over 2\pi} Q
\end{equation}
where $\kappa$ is the surface gravity of the Killing horizon
and $Q$ is an integral over $\Sigma$.
Later we show that for theories where $L_{(m)}$ does not include the
derivatives
of the metric higher than second order
\begin{equation}\label{i5}
Q=8\pi\int_{\Sigma} \sqrt{\sigma} d^2y
{\partial L_{(m)} \over \partial R_{\mu\nu\lambda\rho}}
p_\mu l_\nu p_\lambda l_\rho~~~.
\end{equation}
Here $R_{\mu\nu\lambda\rho}$ is the Riemann tensor of the
background space-time and $p_\mu$, $l_\mu$ are two unit
mutually orthogonal vectors normal to $\Sigma$.
It follows from (\ref{i5}) that $Q$ is not zero when fields
have couplings with the curvature in the Lagrangian.

Because ${\cal E}_{(m)}$ and ${\cal H}_{(m)}$ are different
they are related to the different physical
properties of the system. According to (\ref{i3}),
${\cal H}_{(m)}$ is associated to the generator of the {\it time
evolution}.
The energy ${\cal E}_{(m)}$ is obtained
from the observable stress-energy tensor and should be a {\it physical
energy} of fields $\phi$ on $\Sigma_t$.  In the case of a
black hole,
the role of ${\cal E}_{(m)}$ as the physical energy becomes evident after
examining the first law of black hole thermodynamics.  In the Einstein
gravity the variation of the mass $M$ of a black hole under small
excitation of matter fields $\phi$ with the energy ${\cal E}_{(m)}$
can be represented
as\footnote{This variational formula can be found, for instance, in
book \cite{FrNo:89}.  We omit in (\ref{i6}) the term proportional to
$T^{\mu\nu}\delta g_{\mu\nu}$ by assuming that it is of the second
order in perturbations.}
\begin{equation}\label{i6}
\delta M={\kappa \over 8\pi G} \delta {\cal A}+{\cal E}_{(m)}~~~,
\end{equation}
where $G$
is the Newton constant and ${\cal A}$ is the surface area of
of the horizon.

\bigskip

Thus, in case of black holes, ${\cal H}_{(m)}$ is related
to the time evolution
while ${\cal E}_{(m)}$ is related to the thermodynamical properties
of the system. For the sake of simplicity
we call ${\cal H}_{(m)}$
and ${\cal E}_{(m)}$ the {\it Hamiltonian} and the {\it energy},
respectively.

\bigskip

It should be noted that the difference between ${\cal E}_{(m)}$ and
${\cal H}_{(m)}$ may be crucial.
For instance, this difference
draw an attention under studying statistical-mechanical
interpretation of the Bekenstein-Hawking entropy
in the models of induced gravity
\cite{FFZ}-\cite{FF:98}.  Non-minimal couplings of the constituent
fields with the curvature are an important feature of these models. As
was shown in \cite{FF:97}, \cite{FF:98}, the Noether charge $Q$, Eq.
(\ref{i5}), is not trivial in such models and appears in the formula
for the Bekenstein-Hawking entropy $S^{BH}$
\begin{equation}\label{i7}
S^{BH}=S^{SM}-Q~~~,
\end{equation}
where $S^{SM}$ is the statistical-mechanical (or entanglement)
entropy of the constituents. According to
Eq. (\ref{i6}), the spectrum of the
black hole mass $M$ can be related to the spectrum of
the energy ${\cal E}_{(m)}$
of the constituents. On the other hand, to calculate
$S^{SM}$ one uses the spectrum of
the Hamiltonian ${\cal H}_{(m)}$.
These two facts were used in \cite{FF:98} to
explain the subtraction of $Q$ in Eq. (\ref{i7}).

\bigskip

The present paper  has two aims. The first one is to carry out
a general analysis of
the energy and the Hamiltonian
in diffeomorphism invariant theories on an external
black hole background and derive an
expression for $Q$ (which is reduced to Eq. (\ref{i5})
in the particular case). This also implies studying
stationary geometries with axial symmetry and corresponding
different definitions of the angular momentum of the system.
Our second aim is to demonstrate that
${\cal H}_{(m)}$ and ${\cal E}_{(m)}$ play different roles
and prove an analog of variational formula (\ref{i6})
for black holes in arbitrary diffeomorphism invariant theories of
gravity.

It should be noted that $Q$ is closely related to the Noether
charge which was introduced
by Wald \cite{Wald:93} for description of the
black hole entropy and
studied in Refs. \cite{IW:94}--\cite{Brown:95}.
In many respects our consideration will be parallel to
that of work by Iyer and Wald \cite{IW:94}.

The paper is organized as follows. Section 2 outlines
the Noether charge construction \cite{Wald:93},\cite{IW:94}.
The definition of ${\cal E}_{(m)}$ and ${\cal H}_{(m)}$
and a general form of $Q$ are given in Section 3.
In Section 4 we prove that our definition of ${\cal H}_{(m)}$
gives the generator of time translations.
Then, in Section 5 we obtain a generalization of variational formula
(\ref{i6}) for the energy ${\cal E}_{(m)}$. Our comments regarding
axisymmetric space-times and an extension of the results to rotating
black holes can be found in Section 6. We finish with a
summary and a brief discussion in
Section 7.

\section{Review of the Noether charge and black hole entropy}
\setcounter{equation}0

We begin with a diffeomorphism invariant theory of gravity
which includes matter fields $\phi$.
It is assumed that $\phi$ may have tensor and spinor indices.
We suppose that the system is defined on a space-time ${\cal M}$
with a time-like boundary $\partial {\cal M}$.
The matter fields are assumed to vanish at the past and future
infinities.
The dynamical
equations of the theory are determined by the action
\begin{equation}\label{2.1}
I[g,\phi]=\int_{\cal M} \sqrt{-g} d^4x L[g,\phi]-
\int_{\partial {\cal M}} \sqrt{-h} d^3x B[g,\phi]~~~.
\end{equation}
Here $L[g,\phi]$ is the Lagrangian of the theory which can
be represented as a function of metric $g_{\mu\nu}$,
Riemann tensor $R_{\mu\nu\lambda\rho}$, matter fields $\phi$
and symmetrized covariant derivatives of $R_{\mu\nu\lambda\rho}$
and $\phi$. (For the proof see Ref. \cite{IW:94}.)
$h$ is the determinant of the metric induced on
$\partial {\cal M}$.
The variation of the action has the form \cite{IW:94}
\begin{equation}\label{2.2}
\delta I[g,\phi]=\int_{\cal M}\sqrt{-g} d^4x \left[E^{\mu\nu}
\delta g_{\mu\nu} +E_\phi \delta \phi+\nabla_{\mu}
\theta^\mu(\delta g,\delta \phi)\right]-
\delta\left[\int_{\partial {\cal M}}
\sqrt{-h} d^3x B[g,\phi]\right]~~~.
\end{equation}
Quantities $E^{\mu\nu}$ and $E_\phi$ depend on the background
fields $g_{\mu\nu}$ and $\phi$
only. The components
$\theta^\mu$ of a one-form
depend on the background fields, variations $\delta\phi$,
$\delta g$ and their covariant derivatives.  To have
a well-defined variational procedure it is assumed that
the following equality
\begin{equation}\label{2.3}
\int_{\partial {\cal M}}\sqrt{-h} d^3x
u_\mu \theta^\mu(\delta g,\delta \phi)=-\delta
\left[\int_{\partial {\cal M}}\sqrt{-h} d^3x
B[g,\phi]\right]~~~
\end{equation}
takes place for given boundary
conditions\footnote{Note that Eq. (\ref{2.3}) is necessary but
not sufficient condition to fix the functional $B$.}.
Here $u_{\mu}$ is a unit inward-pointing vector normal to
the boundary $\partial {\cal M}$.
The action has an
extremum when $g_{\mu\nu}$ and $\phi$ obey the equations
\begin{equation}\label{2.4}
E^{\mu\nu}=0~~~,
\end{equation}
\begin{equation}\label{2.5}
E_\phi=0~~~.
\end{equation}
Equation (\ref{2.5}) is the equation of motion of the matter fields.
Consider now an infinitesimal transformation of coordinates and fields
\begin{equation}\label{2.6}
x'^\mu=x^\mu-\xi^\mu(x)~~~,
\end{equation}
\begin{equation}\label{2.7}
\phi'(x)=\phi(x)+{\cal L}_\xi\phi(x)~~~,
\end{equation}
\begin{equation}\label{2.8}
g'_{\mu\nu}(x)=g_{\mu\nu}(x)+{\cal L}_\xi g_{\mu\nu}(x)~~~,
\end{equation}
where ${\cal L}_\xi$ is the Lie derivative along the vector field
$\xi^\mu$.
When $\xi^\mu(x)$ has a compact support,
it
follows from (\ref{2.2}), (\ref{2.6})--(\ref{2.8}) that
\begin{equation}\label{2.9}
\delta_\xi I=\int_{\cal M}\sqrt{-g} d^4x \left[E^{\mu\nu}\
{\cal L}_\xi g_{\mu\nu}+E_\phi {\cal L}_\xi \phi+\nabla_\mu
(\theta^\mu(\xi)-\xi^\mu L)\right]+O(\xi^2)~~~,
\end{equation}
where
\begin{equation}\label{2.10}
\theta^\mu(\xi)\equiv \theta^\mu({\cal L}_\xi g,{\cal L}_\xi \phi)~~~.
\end{equation}
In a diffeomorphism invariant theory $\delta_\xi I=0$ and
each coordinate transformation generates
a Noether current
\begin{equation}\label{2.11}
J^\mu(\xi)=\theta^\mu(\xi)-\xi^\mu L
\end{equation}
which conserves
\begin{equation}\label{2.12}
\nabla^\mu J_\mu(\xi)=0
\end{equation}
provided if equations of motion (\ref{2.4}), (\ref{2.5})
are satisfied.
By taking into account that (\ref{2.12}) holds for any $\xi^\mu(x)$
one can prove \cite{Wald} that there is such a tensor
$Q^{\mu\nu}(\xi)=-Q^{\nu\mu}(\xi)$ that
\begin{equation}\label{2.13}
J^\mu(\xi)=\nabla_\nu Q^{\mu\nu}(\xi)~~~.
\end{equation}
The quantity $Q^{\mu\nu}(\xi)$ is called the {\it Noether potential}.
The integral over a two-dimensional surface $\Sigma$
\begin{equation}\label{2.14}
Q(\Sigma,\xi)=c\int_{\Sigma} Q^{\mu\nu}(\xi)d\sigma_{\mu\nu}
\end{equation}
is called the {\it Noether charge}. Here $c$ is a normalization
constant which will be fixed later.

Suppose now that ${\cal M}$ is a stationary asymptotically flat
black hole background. Denote by $\xi^\mu$ a Killing vector
\begin{equation}\label{2.15}
\xi^\mu=t^\mu+\Omega_H \varphi^\mu~~~,
\end{equation}
where $t^\mu$ is a time-like Killing vector corresponding to
time translations of the system and $\varphi^\mu$ is
a Killing vector corresponding to rotations. The coefficient
$\Omega_H$ coincides with the angular velocity near the
horizon where $\xi^2=0$.
The bifurcation surface $\Sigma$ is determined by
the condition $\xi^\mu=0$.

The total energy ${M}$
and angular momentum $\cal J$ of the system conserve.
On equations of motion (\ref{2.4}) and (\ref{2.5})  these quantities
are reduced to surface integrals on a two-dimensional spatial
boundary $C_t$ of $\Sigma_t$ ($C_t$ is the
intersection of $\Sigma_t$ and $\partial {\cal
M}$). According to \cite{IW:94},
\begin{equation}\label{2.16}
{M}=\int_{C_t}\sqrt{\gamma}d^2y (Q^{\mu\nu}(t)n_\mu u_\nu
+B N)~~~,
\end{equation}
\begin{equation}\label{2.17}
{\cal J}=-\int_{C_t} \sqrt{\gamma}d^2y Q^{\mu\nu}(\varphi) n_\mu
u_\nu~~~.
\end{equation}
Here $N=\sqrt{-g_{00}}$, $\gamma$ is the
metric induced on $C_t$, and $n^\mu$ is the future-directed unit vector
normal to $\Sigma_t$.  When $\Sigma_t$ is an infinite surface we
will write $C_\infty$ instead of $C_t$ for the boundary at asymptotic
infinity.  By following \cite{IW:94} we
assume that $\varphi^\mu$ is everywhere tangent at  $C_\infty$.  One
can show \cite{IW:94} that in the Einstein gravity the boundary
function $B$ can be chosen so that $M$, calculated at $C_\infty$,
coincides  with the ADM mass of the black hole.

Hypersurfaces of constant time $t$
intersect at the bifurcation surface $\Sigma$ of the
Killing horizons.
$\Sigma$ is an inner boundary of $\Sigma_t$.
Let us define the following quantity
\begin{equation}\label{2.18}
S={2\pi \over \kappa} \int_{\Sigma} \sqrt{\sigma}d^2z
Q^{\mu\nu}(\xi) l_\mu p_\nu
\end{equation}
where $\kappa$ is the surface gravity of the horizon
and $l_\nu$ and $p_\mu$  are unit mutually orthogonal vectors
normal to $\Sigma$. We assume that
$l^\mu$ is
inward-pointing and $p^\mu$ is future-directed vectors
normalized as $l^2=-p^2=1$.
The important result by Wald \cite{Wald:93} relates the variation of the
integral (\ref{2.18}) on $\Sigma$ to the variation of the total energy
and momentum computed on $C_t$. Namely,
\begin{equation}\label{2.19}
\delta M=T_H \delta S+\Omega_H
\delta {\cal J}~~~,
\end{equation}
where $T_H=\kappa/(2\pi)$. This
equation can be interpreted as a first law of black hole thermodynamics
(mechanics) in a general diffeomorphism invariant theory of gravity.
The quantity $T_H$ is associated to the temperature of a black hole,
and $S$ plays the role of the black hole entropy.
Formula (\ref{2.19}) holds when the background fields are the
solutions of Eqs. (\ref{2.4}), (\ref{2.5}) while variations
$\delta g$ and $\delta \phi$ obey the linearized equations
of motion. Note that it is not required that $\delta g$ and $\delta\phi$
preserve the symmetries of the background solutions (i.e., that
their Lie derivatives along $\xi^\mu$ vanish).

The explicit form of $S$ is determined the general structure of
the Noether potential found out by Iyer and Wald \cite{IW:94}
\begin{equation}\label{2.20}
Q^{\mu\nu}(\xi)=2E^{\mu\nu\lambda\rho}\nabla_\lambda\xi_{\rho}
+W^{\mu\nu\lambda}\xi_\lambda~~~,
\end{equation}
where the quantities $E^{\mu\nu\lambda\rho}$ and $W^{\mu\nu\lambda}$
do not depend on $\xi^\mu$. The second term in (\ref{2.20}) does not
contribute to $S$ because $\xi^\mu=0$ on $\Sigma$, and one can show that
\begin{equation}\label{2.21}
S=-8\pi\int_{\Sigma}\sqrt{\sigma} d^2z E^{\mu\nu\lambda\rho}
p_\mu l_\nu p_\lambda l_\rho~~~.
\end{equation}
The quantity $E^{\mu\nu\lambda\rho}$ is \cite{IW:94}
\begin{equation}\label{2.22}
E^{\mu\nu\lambda\rho}=\hat{X}^{\mu\nu\lambda\rho} L~~~,
\end{equation}
\begin{equation}\label{2.23}
\hat{X}^{\mu\nu\lambda\rho}\equiv {\partial \over \partial
R_{\mu\nu\lambda \rho}}-\nabla_{\gamma_1}{\partial \over
\partial \nabla_{\gamma_1} R_{\mu\nu\lambda\rho}}
+...+(-1)^m \nabla_{(\gamma_1...}\nabla_{\gamma_m)}
{\partial \over \partial \nabla_{(\gamma_1...}\nabla_{\gamma_m)}
R_{\mu\nu\lambda\rho}}~~~.
\end{equation}
Here $m$ is the highest derivative of the Riemann tensor in the
Lagrangian $L$ of the theory, and symbol $\nabla_{(\gamma_1...}
\nabla_{\gamma_m)}$ denotes symmetrization of the covariant
derivatives. The partial derivatives in (\ref{2.23}) are
uniquely fixed by requiring them to have the same tensor
symmetries as varied quantities.

It should be noted that there is a freedom in the definition
of the Noether potential $Q^{\lambda\rho}$. First, one can add to
the Lagrangian $L$ a total derivative $\nabla\mu$
\footnote{This addition does not change equations of motion
(\ref{2.4}), (\ref{2.5}). However, it is equivalent to a modification
of the boundary functional $B$ by the term $u^\lambda \mu_\lambda$ on
$\partial {\cal M}$. This modification must not violate the
variational procedure of the action $I$, otherwise it results in
a change of the boundary conditions imposed on the fields.}. As a
result, $Q^{\lambda\rho}$ changes as
\begin{equation}\label{2.24} \Delta
Q^{\lambda\rho}(\xi)=\mu^\lambda \xi^\rho-\mu^\rho\xi^\lambda~~~.
\end{equation}
Second, Eq. (\ref{2.9}) does not determine the Noether current
(\ref{2.11}) uniquely.
One can add to $J^\mu(\xi)$ the term
$\nabla_\lambda Y^{\mu\lambda}$, where $Y^{\mu\lambda}=-Y^{\mu\lambda}$
is linear in the varied fields. This results in a change of the
potential
\begin{equation}\label{2.25}
\Delta Q^{\lambda\rho}(\xi)=Q^{\lambda\rho}(g,\phi,{\cal L}_\xi g,
{\cal L}_\xi \phi)~~~.
\end{equation}
Finally, Eq. (\ref{2.13}) allows  the freedom to add the term
\begin{equation}\label{2.26}
\Delta Q^{\lambda\rho}(\xi)=\nabla_\nu Z^{\lambda\rho\nu}~~~,
\end{equation}
where $Z^{\lambda\rho\nu}$ is a totally antisymmetric tensor.
It is important that the black hole entropy $S$ is defined on the
bifurcation surface $\Sigma$ where $\xi^\mu=0$ and so
adding terms (\ref{2.24}) and (\ref{2.25}) to the potential
does not change $S$. Regarding term (\ref{2.26}), it vanishes in $S$
because $\Sigma$ is a closed surface.

\section{Energy and Hamiltonian of matter fields}
\setcounter{equation}0

Let us define now the energy and
the Hamiltonian of matter fields on an external gravitational
background. We start with the total action (\ref{2.1}) and
split it onto two parts
\begin{equation}\label{3.1}
I[g,\phi]=I_{(g)}[g]+I_{(m)}[g,\phi]~~~,
\end{equation}
\begin{equation}\label{3.2}
I_{(g)}[g]=\int_{\cal M} \sqrt{-g} d^4x L_{(g)}[g]-
\int_{\partial {\cal M}} \sqrt{-h} d^3x
B_{(g)}[g]~~~,
\end{equation}
\begin{equation}\label{3.3}
I_{(m)}[g,\phi]=\int_{\cal M} \sqrt{-g}d^4x L_{(m)}[g,\phi]-
\int_{\partial {\cal M}}\sqrt{-h}d^3 x B_{(m)}[g,\phi]~~~.
\end{equation}
Functional $I_{(g)}$ represents a pure gravitational action without
the matter ("in vacuum"), while $I_{(m)}$ is the action of matter fields
$\phi$ in an external gravitational field $g_{\mu\nu}$.
Without loss
of generality we assume that
\begin{equation}\label{3.4}
I_{(m)}[g,\phi=0]=0~~~.
\end{equation}
According to Eq. (\ref{3.1}), the form $\theta^\mu(\xi)$,
introduced in (\ref{2.2}),
can be written as
\begin{equation}\label{3.1a}
\theta^\mu(\xi)=\theta^\mu_{(g)}(\xi)+\theta^\mu_{(m)}(\xi)~~~.
\end{equation}
The quantities $\theta^\mu_{(g)}(\xi)$ and $\theta^\mu_{(m)}(\xi)$
are determined by
the higher order derivatives of $g_{\mu\nu}$, $\phi$ in the
Lagrangians $L_{(g)}$ and $L_{(m)}$,
respectively.
$\theta^\mu_{(g)}(\xi)$ and $\theta^\mu_{(m)}(\xi)$
are linear in variations ${\cal L}_\xi g$ and
${\cal L}_\xi\phi$ .

Note that equations of motion (\ref{2.5}) of the fields $\phi$ are
determined by the variation of the matter action only,
\begin{equation}\label{3.7}
{1 \over \sqrt{-g}}
{\delta I_{(m)} \over \delta \phi}=E_\phi=0~~~.
\end{equation}
Consider now diffeomorphism
transformations (\ref{2.6})--(\ref{2.8}) in the matter action. By
assuming that the background fields $\phi$ obey (\ref{3.7}) and
their variations have a compact support one finds
\begin{equation}\label{3.5}
\delta_\xi I_{(m)}=\int_{\cal M}\sqrt{-g}d^4x \left[\frac 12 T^{\mu\nu}
{\cal L}_\xi g_{\mu\nu}+\nabla_\mu(\theta_{(m)}^\mu(\xi)-\xi^\mu L_{(m)})
\right]~~~.
\end{equation}
Here we introduced the stress-energy tensor of the matter
\begin{equation}\label{3.6}
T^{\mu\nu}={2 \over \sqrt{-g}} {\delta I_{(m)} \over \delta g_{\mu\nu}}~~~.
\end{equation}
In a diffeomorphism invariant
theory $\delta_\xi I_{(m)}=0$ and
we find from (\ref{3.5})
\begin{equation}\label{3.8}
\nabla_\mu(\theta_{(m)}^\mu(\xi)-\xi^\mu L_{(m)}+\xi_\nu T^{\mu\nu})-\xi_\nu
T^{\nu\mu}_{~~~;\mu}=0~~~.
\end{equation}
On equations of motion (\ref{3.7}) the divergence of $T^{\mu\nu}$
vanishes\footnote{Indeed, it follows from (\ref{3.8}) that for any
$\xi^\mu$ the quantity $T^{\mu\nu}_{~~~;\mu}\xi_\nu$ should be
the divergence of a vector. It is possible only when (\ref{3.9}) holds.}
\begin{equation}\label{3.9}
T^{\mu\nu}_{~~~;\nu}=0~~~.
\end{equation}
Thus, there is a conservation law
\begin{equation}\label{3.10}
\nabla_\mu J_{(m)}^\mu(\xi)=0~~~,
\end{equation}
for the following vector
\begin{equation}\label{3.11}
J_{(m)}^\mu(\xi)=-\theta_{(m)}^\mu(\xi)+\xi^\mu L_{(m)}-T^{\mu\nu}\xi_\nu~~~.
\end{equation}
Let us emphasize that for Eq. (\ref{3.10}) to hold only equations
of matter fields (\ref{3.7}) are required while the background
metric $g_{\mu\nu}$ can be arbitrary. Thus, $J_{(m)}^\mu(\xi)$ is the
Noether current of the matter in an external
gravitational field.

\bigskip

Suppose now that the background space ${\cal M}$ is
invariant
with respect to time translations defined by the
Killing vector field $t^\mu$.
Let $\Sigma_t$ be the surface
of the constant time $t$ and define on this surface the
quantities
\begin{equation}\label{3.13}
{\cal H}_{(m)}=-\int_{\Sigma_t} (\theta_{(m)}^\mu(t)-t^\mu L_{(m)})d\Sigma_\mu~~~,
\end{equation}
\begin{equation}\label{3.14}
{\cal E}_{(m)}=\int_{\Sigma_t}T^{\mu\nu}t_\mu d\Sigma_\nu~~~,
\end{equation}
where $d\Sigma^\mu$ is the future-directed vector of the volume element
on $\Sigma_t$ \footnote{$d\Sigma^\mu=n^\mu\tilde{h}^{1/2}d^3x$ where
$\tilde{h}$ is the Jacobian of the metric on $\Sigma_t$
and $n^\mu$ is the unit vector orthogonal to $\Sigma_t$,
$n_\mu=(-N,0,0,0)$.}.
We call the functionals ${\cal H}_{(m)}$ and ${\cal E}_{(m)}$
the {\it Hamiltonian} and the
{\it energy}, respectively.

Let us note that the forms $\theta^\mu_{(m)}(t)$
depend on the variations of the fields $\phi$ only. Variations
of the background metric vanish ${\cal L}_t g_{\mu\nu}=0$.
It is easy to see that definition of ${\cal E}_{(m)}$ coincides
with standard formula (\ref{i1}). To justify the definition of
the Hamiltonian let us consider a theory with non-degenerate Lagrangian
$L_{(m)}$ which does not contain the derivatives of $\phi$ higher than
the first order.
In such theories
\begin{equation}\label{3.15}
n_\mu \theta_m^\mu(t)=n_\mu{\partial L_{(m)} \over \partial \nabla_\mu \phi}
{\cal L}_t\phi=-N {\partial L_{(m)} \over \partial \dot{\phi}} {\cal L}_t\phi~~~,
\end{equation}
where $\dot{\phi}=d\phi/dt$.
By taking into account (\ref{3.15}) and the identities $N\tilde{h}^{1/2}=
\sqrt{-g}$, $t^\mu n_\mu=-N$ one
obtains ${\cal H}_{(m)}$ in the standard form (\ref{i3}).
In the next Section we will
show explicitly that ${\cal H}_{(m)}$ corresponds
to the generator of canonical evolution of the matter fields
along the Killing time $t$.

It is worth considering the following "currents"
\begin{equation}\label{3.16}
j_{(m)}^\mu(\xi)=T^{\mu\nu}\xi_\nu~~~,~~~
q_{(m)}^\mu(\xi)=-\theta_{(m)}^\mu(\xi)+\xi^\mu L~~~.
\end{equation}
When $\xi^\mu=t^\mu$ one can verify the conservation laws
\begin{equation}\label{3.17}
\nabla_\mu j_{(m)}^\mu(t)=0~~~,~~~\nabla_\mu q_{(m)}^\mu(t)=0~~~,
\end{equation}
which follow from equations (\ref{3.9})--(\ref{3.11}).
Equations (\ref{3.17}) imply that the energy ${\cal E}_{(m)}$
and the Hamiltonian ${\cal H}_{(m)}$ do not change when the surface
$\Sigma_t$ undergoes variations of a compact support.
Moreover, if the boundary terms $\int j_{(m)}^\mu(t) dB_\mu$ and
$\int q_{(m)}^\mu(t)dB_\mu$ on $\partial {\cal M}$ are zero\footnote{Here
$dB^\mu$ is the vector of the volume element
of the boundary surface $\partial {\cal M}$. The boundary terms
vanish, for instance, when $\partial {\cal M}$ is at
asymptotic infinity and
the fields decay fast enough in this region.}
${\cal E}_{(m)}$ and ${\cal H}_{(m)}$ do not depend on the time
parameter $t$ which labels the foliation $\Sigma_t$.

\bigskip

Let us find now the relation between the Hamiltonian and the
energy.
According to Eq. (\ref{3.11}),
\begin{equation}\label{3.12}
{\cal H}_{(m)}-{\cal E}_{(m)}=
\int_{\Sigma_t} J_{(m)}^\mu(t)d\Sigma_\mu~~~.
\end{equation}
One can introduce
a Noether potential $Q_{(m)}^{\mu\nu}$ corresponding to the
Noether current $J_{(m)}^\mu$.
For any $\xi^\mu$,
\begin{equation}\label{3.18}
J_{(m)}^\mu(\xi)=\nabla_\nu Q_{(m)}^{\mu\nu}(\xi)~~~.
\end{equation}
Then Eq. (\ref{3.12}) can be rewritten as
\begin{equation}\label{3.19}
{\cal H}_{(m)}-{\cal E}_{(m)}=
\int_{C_t} \sqrt{\gamma} d^2y
Q_{(m)}^{\mu\nu}(t)u_\mu n_\nu~~~.
\end{equation}
Here $C_t$ is the boundary of $\Sigma_t$, i.e., the intersection
of $\Sigma_t$ and $\partial {\cal M}$. As in Eqs. (\ref{2.16}),
(\ref{2.17}),
$u_\mu$ is a unit
inward normal of $\partial {\cal M}$ and $n_\mu$ is future-directed
normal of $\Sigma_t$.
(Both vectors are unit mutually othogonal normals of
surface $C_t$).  The quantity in r.h.s. of (\ref{3.19})  is a Noether
charge defined with respect to the surface $C_t$ and the Killing vector
$t^\mu$.
Therefore, we proved that the difference of the energy and the
Hamiltonian is a surface term. In many physical situations
${\cal M}$ has the topology $R\times \Sigma_t$ and the surface
$C_t$ is chosen at the spatial infinity. If the fields rapidly
decay at infinity the boundary term in r.h.s of (\ref{3.19})
vanishes and ${\cal H}_{(m)}$ and ${\cal E}_{(m)}$ coincide.

Consider the case of black hole space-times. In this instance
we assume that the space-time
is static and the Killing vector $t^\mu=0$ on the
bifurcation surface $\Sigma$. All surfaces $\Sigma_t$ of the
foliation meet at $\Sigma$. Although $\Sigma$ plays a role of the
inner boundary of $\Sigma_t$ the fields subject
no conditions on this surface except regularity.
The region of $\cal M$ between $\Sigma$ and $\partial {\cal M}$ is
an exterior region of a black hole which has the topology
$R^2\times\Sigma$. For the black hole exterior relation
(\ref{3.19})  can be written as
\begin{equation}\label{3.20}
{\cal H}_{(m)}-{\cal E}_{(m)}=
\int_{\Sigma} \sqrt{\sigma}d^2z
Q_{(m)}^{\mu\nu}(t)l_\mu p_\nu
+\int_{C_t} \sqrt{\gamma}d^2y
Q_{(m)}^{\mu\nu}(t) u_\mu n_\nu ~~~,
\end{equation}
where $l^\mu$ and $p^\mu$ are normals of $\Sigma$ defined
as in Eq. (\ref{2.18}).
The last term in r.h.s. disappears when fields vanish on $C_t$
and one gets
\begin{equation}\label{3.21}
{\cal H}_{(m)}-{\cal E}_{(m)}=
\int_{\Sigma} \sqrt{\sigma} d^2z
Q_{(m)}^{\mu\nu}(t)l_\mu p_\nu ~~~.
\end{equation}
Thus, we can conclude that the energy and the Hamiltonian
are always different for the theories where the Noether potential
$Q_{(m)}^{\mu\nu}(t)$
at the bifurcation surface is not zero.

In fact,
an explicit expression for the Noether charge in r.h.s. of
(\ref{3.21}) follows from a consideration similar to that
of Section 2. First, let us note that
the only difference between the forms of
the Noether current for matter fields $J_{(m)}^\mu(\xi)$, Eq. (\ref{3.11}),
and the current of the complete theory $J^\mu(\xi)$, Eq. (\ref{2.11}),
is the term $-T^{\mu\nu}\xi_\nu$ in $J_{(m)}^\mu(\xi)$.
The matter current does not include the derivatives of the
vector $\xi^\mu$ higher than the third order. Thus, by taking into
account Eq. (\ref{3.18}) one can write\footnote{Note that potentials
$Q_{(m)}^{\mu\nu}$ and $Q^{\mu\nu}$ have different signs.
Such a difference is a matter of our convention of the definition
of $J_{(m)}^\mu$. We follow Refs. \cite{FF:97},\cite{FF:98} where
this convention was used first.}
\begin{equation}\label{3.22}
Q_{(m)}^{\mu\nu}(\xi)=-2E_{(m)}^{\mu\nu\lambda\rho}\nabla_\lambda
\xi_{\rho}
-W_{(m)}^{\mu\nu\lambda}\xi_\lambda~~~,
\end{equation}
where quantities $E_{(m)}^{\mu\nu\lambda\rho}$ and
$W_{(m)}^{\mu\nu\lambda}$ depend on the background fields only. Here
$Q_{(m)}^{\mu\nu}$ is presented in the same form as  $Q^{\mu\nu}$, see
(\ref{2.20}).  Because the second term in r.h.s. of (\ref{3.22})
vanishes on $\Sigma$ we find from (\ref{3.21})
\begin{equation}\label{3.23}
{\cal H}_{(m)}-{\cal E}_{(m)}={\kappa \over 2\pi} Q_{(m)}~~~,
\end{equation}
\begin{equation}\label{3.24}
Q_{(m)}=8\pi\int_{\Sigma}\sqrt{\sigma} d^2z
E_{(m)}^{\mu\nu\lambda\rho} p_\mu l_\nu
p_\lambda l_\rho
\end{equation}
To obtain Eqs. (\ref{3.23}) and (\ref{3.24}) we took into account
that $t_{\mu;\nu}=\kappa(p_\mu l_\nu-p_\nu l_\mu)$ on $\Sigma$,
where $\kappa$ is the surface gravity of the Killing horizon.
Obviously, the tensor $E_{(m)}^{\mu\nu\lambda\rho}$ does not
depend on the presence of the term
$-T^{\mu\nu}\xi_\nu$ in $J_{(m)}^\mu$.
This implies that
$E_{(m)}^{\mu\nu\lambda\rho}$
has precisely the same form
as the tensor $E^{\mu\nu\lambda\rho}$ in $Q^{\mu\nu}(\xi)$.
Thus, as follows from (\ref{2.22}),
\begin{equation}\label{3.25}
E_{(m)}^{\mu\nu\lambda\rho}=\hat{X}^{\mu\nu\lambda\rho} L_{(m)}~~~.
\end{equation}
$L_{(m)}$ is the Lagrangian of matter fields, see (\ref{3.3}),
and the operator $\hat{X}^{\mu\nu\lambda\rho}$ is
defined by (\ref{2.23}).

According to equations (\ref{3.23})--(\ref{3.25}), a {\it necessary}
condition for the energy to differ from the Hamiltonian
is the presence of {\it non-minimal couplings} of matter fields with
the curvature of the space-time.
To put it in another way,
$L_{(m)}$
has to depend on $R_{\mu\nu\lambda\rho}$ and its derivatives.
In the case when $L_{(m)}$ does not include the derivatives
of $R_{\mu\nu\lambda\rho}$
the Noether charge
$Q_{(m)}$ takes simple form (\ref{i5}).

\bigskip

By comparing Eqs. (\ref{2.21}), (\ref{2.22})
with (\ref{3.24}), (\ref{3.25}) one makes another observation:
the charge $Q_{(m)}$ is the contribution of the matter fields
to the entropy of a black hole. Indeed, it follows from
the decomposition of the total action onto gravitational
and matter parts, Eqs. (\ref{3.1})--(\ref{3.3}), that
the black hole entropy is represented as
\begin{equation}\label{3.26}
S=S_{(g)}+S_{(m)}~~~,
\end{equation}
\begin{equation}\label{3.27}
S_{(m)}=-Q_{(m)}~~~.
\end{equation}
The term $S_{(g)}$ is the pure gravitational part of the entropy
determined by the Lagrangian $L_{(g)}$
\begin{equation}\label{2.21a}
S_{(g)}=-8\pi\int_{\Sigma}\sqrt{\sigma} d^2z E_{(g)}^{\mu\nu\lambda\rho}
p_\mu l_\nu p_\lambda l_\rho~~~,
\end{equation}
\begin{equation}\label{2.22a}
E_{(g)}^{\mu\nu\lambda\rho}=\hat{X}^{\mu\nu\lambda\rho} L_{(g)}~~~.
\end{equation}

\bigskip

Also, it should be noted that the Noether potential
$Q_{(m)}^{\mu\nu}$
is not determined uniquely. It changes, for instance, when
one adds a total divergence
to the matter Lagrangian $L_{(m)}$.
Possible changes of $Q_{(m)}^{\mu\nu}$ are
analogous to changes of $Q^{\mu\nu}$ and are described by
Eqs. (\ref{2.24})--(\ref{2.26}). It is important, however,
that the charge $Q_{(m)}$ is defined on the
bifurcation surface $\Sigma$ and does not depend on this freedom.

Finally, let us emphasize that in the above analysis we assumed that
the space-time is static.
A generalization of these
results  to stationary
axisymmetric spacetimes including rotating black holes
will be given in Section 6.

\section{${\cal H}_{(m)}$ as a generator of canonical transformations}
\setcounter{equation}0

In what follows we study properties of the Hamiltonian
${\cal H}_{(m)}$ and the energy ${\cal E}_{(m)}$. In this section we
prove that ${\cal H}_{(m)}$ is the generator of the canonical
transformation of the system along the Killing time $t$.
To define generators of canonical transformations
of a field system on a curved background we
choose the formalism developed by DeWitt in \cite{DeWitt:65}.

Let us first assume that ${\cal M}$ has the topology $R\times\Sigma_t$.
We also suppose that fields vanish fast enough at
$t\rightarrow \pm \infty$ and consider
"observables" which are functionals of $\phi$ on
$\cal M$.  The Poisson bracket of two "observables" represented by
functionals $A[\phi]$ and $B[\phi]$
can be defined as \cite{Pierls:52}, \cite{DeWitt:65}
\begin{equation}\label{4.1}
(A,B)=\int d^4x d^4y {\delta A \over \delta \phi(x)}
\tilde{G}(x,y){\delta B \over \delta\phi(y)}~~~.
\end{equation}
Here $\tilde{G}$ is the Pauli-Jordan function of the theory
which is expressed in terms of the advanced, $G^{+}$, and
retarded, $G^{-}$, Green functions
\begin{equation}\label{4.2}
\tilde{G}(x,y)=G^{+}(x,y)-G^{-}(x,y)~~~.
\end{equation}
Functions $G^{\pm}$ obey the equation
\begin{equation}\label{4.3}
\int d^4y
{\delta^2 I_{(m)}[g,\phi] \over \delta \phi(x)\delta
\phi(y)}G^{\pm}(y,z)= -\delta^{(4)}(x-z)~~~,
\end{equation}
and
additional conditions which in local theories are
\begin{equation}\label{4.5}
G^{+}(x,y)=0~~~,~~~t_x >t_y~~~,
\end{equation}
\begin{equation}\label{4.6}
G^{-}(x,y)=0~~~,~~~t_x<t_y~~~.
\end{equation}
where $t_x$, $t_y$ are the time coordinates of the points $x$
and $y$, respectively.
We assume that
$\delta^2 I_{(m)} /\delta \phi^2$ is a non-degenerate
operator and $G^{\pm}$
can be uniquely defined by Eq. (\ref{4.3}) under chosen boundary
conditions\footnote{Our analysis can be also carried out for degenerate
operators which appear in gauge theories. The problem of singular
operators is resolved by imposing gauge conditions \cite{DeWitt:65}.
Because there are no other changes we do not pay here a special
attention to this case.}.  A simple example which illustrates
(\ref{4.3}) is a theory of a free massive non-minimally coupled scalar
field described by the action
\begin{equation}\label{4.7}
I_{(m)}[g,\phi]=
-\frac 12 \int \sqrt{-g}d^4x \left((\nabla\phi)^2+m^2\phi^2
+\xi R\phi^2\right)~~~.
\end{equation}
For this theory equations (\ref{4.3}) are reduced to
\begin{equation}\label{4.8}
(\nabla^\mu\nabla_\mu-m^2-\xi R)G^{\pm}(x,y)
=-(-g(x))^{-1/2}\delta^{(4)}(x-y)~~~.
\end{equation}

As follows from (\ref{4.1}), the Poisson bracket of two fields
$\phi(x)$ and $\phi(y)$ is
\begin{equation}\label{4.9}
(\phi(x),\phi(y))=\tilde{G}(x,y)~~~.
\end{equation}
In quantum theory this relation is replaced by the commutator
(or anticommutator)
of the corresponding operators
\begin{equation}\label{4.10}
[\hat{\phi}(x),\hat{\phi}(y)]=i\tilde{G}(x,y)~~~.
\end{equation}

We demonstrate now that ${\cal H}_{(m)}$, as
defined by Eq. (\ref{3.13}), is the generator of
canonical transformations of the form
$\delta\phi(x)={\cal L}_t\phi(x)$.
Let us fix the hypersurface
$\Sigma_t$ in (\ref{3.13}) at some constant time $t$ and split the
action onto two parts
\begin{equation}\label{4.11}
I_{(m)}[g,\phi]=I_{(m)}^{+}[g,\phi]+I_{(m)}^{-}[g,\phi]~~~.
\end{equation}
Functionals $I_{(m)}^{\pm}$ are defined in the regions ${\cal M}^{\pm}$
of ${\cal M}$. Points of ${\cal M}^{+}$ and ${\cal M}^{-}$
are in the future or in the past of
$\Sigma_t$, respectively. We have
\begin{equation}\label{4.12}
I_{(m)}^{\pm}[g,\phi]=\int_{{\cal M}^{\pm}}\sqrt{-g} d^4x
L_{(m)}[g,\phi]-\int_{\partial {\cal M}^{\pm}} \sqrt{-h}d^3x
B_{(m)}[g,\phi]~~~.
\end{equation}
Note that ${\cal M}^{\pm}$ have two
boundaries:  the time-like boundary $\partial {\cal M}^{\pm}$ and
 $\Sigma_t$. By using diffeomorphism invariance of the action
one obtains the equality
$$
\delta_t I_{(m)}^{\pm}
=\int_{{\cal M}^{\pm}}\sqrt{-g} d^4x
\left[E_\phi {\cal L}_t\phi+\nabla_\mu
\left(\theta_{(m)}^\mu(t)-t^\mu L_{(m)}\right)\right]
$$
\begin{equation}\label{4.13}
-\delta_t
\left[\int_{\partial {\cal M}^{\pm}} \sqrt{-h} d^3x B_{(m)}\right]=0~~~.
\end{equation}
As a result of the definition of $B_{(m)}$, see Eq. (\ref{2.3}),
\begin{equation}\label{4.14}
\int_{\partial {\cal M}^\pm}\sqrt{-h} d^3x u_\mu
\theta_{(m)}^\mu(t)=-\delta_t\left[
\int_{\partial {\cal M}^\pm}\sqrt{-h} d^3x
B_{(m)}\right]~~~
\end{equation}
and the boundary terms on $\partial {\cal M}^{\pm}$ are canceled.
From (\ref{4.13}) we find that
\begin{equation}\label{4.15}
\int_{{\cal M}^{\pm}}\sqrt{-g} d^4x E_\phi {\cal L}_t\phi=\pm
{\cal H}_{(m)}~~~.
\end{equation}
Let us make a
variation
of the both parts of (\ref{4.15}) over the field
$\phi$
and replace afterwords the
background field $\phi$ by a solution of classical equations
(\ref{3.7}). One gets
\begin{equation}\label{4.16}
\int_{{\cal M}^{\pm}} d^4y
{\delta^2 I_{(m)} \over \delta \phi(x) \delta \phi(y)} {\cal
L}_t\phi(y)=\pm {\delta {\cal H}_{(m)} \over \delta \phi(x)} ~~~.
\end{equation}
We can consider (\ref{4.16}) as an equation which defines
${\cal L}_t \phi$. By using definition (\ref{4.3}) and conditions
(\ref{4.5}), (\ref{4.6})  we find
\begin{equation}\label{4.17}
{\cal L}_t\phi(x)=\mp \int_{{\cal M}^{\pm}}
d^4y G^{\mp}(x,y) {\delta {\cal H}_{(m)}
\over \delta \phi(y)}~~~,
\end{equation}
where the signs "$\mp$" correspond to the field
${\cal L}_t\phi$ in the regions
${\cal M}^{\pm}$. Finally, by taking into account (\ref{4.2})
we have
\begin{equation}\label{4.18}
{\cal L}_t\phi(x)=(\phi(x),{\cal H}_{(m)})~~~,
\end{equation}
where the Poisson bracket is defined by (\ref{4.1}).
As a consequence, variation with time of any "observable"
$A$ is generated by ${\cal H}_{(m)}$
\begin{equation}\label{4.19}
\delta_t A=\int_{\cal M}d^4x {\delta A \over \delta\phi(x)}
{\cal L}_t\phi(x)=(A,{\cal H}_{(m)})~~~.
\end{equation}
Therefore, we demonstrated that ${\cal H}_{(m)}$ coincides
with the generator of the canonical transformations along the
Killing time $t$. As was shown in \cite{DeWitt:65}, the formalism based
on using brackets (\ref{4.1}) is equivalent to the standard
canonical formalism. According to (\ref{4.10}), in a quantized
theory Eq. (\ref{4.18}) becomes
\begin{equation}\label{4.20}
i{\cal L}_t\hat{\phi}(x)=[\hat{\phi}(x),\hat{H}_{(m)}]~~~
\end{equation}
which is the standard relation.

\bigskip

Generalization of this formalism to
black hole space-times requires some care because
the surface $\Sigma_t$ has the internal boundary $\Sigma$
where the Killing vector $t^\mu$ vanishes. To avoid
this complication let us consider the surface $\Sigma^\epsilon_t$
which is obtained from $\Sigma_t$ by cutting the region near
$\Sigma$. We thus assume that $\Sigma^\epsilon_t$ has an
inner boundary which lies near $\Sigma$
at a proper distance $\epsilon$, where $\epsilon$ is a small parameter.
The foliation of the surfaces $\Sigma^\epsilon_t$ with different $t$
represents a region ${\cal M}^\epsilon$ with an inner boundary
${\cal B}^\epsilon$ near the Killing horizon. Consider now
Eq. (\ref{4.13}) in this region. It results in identity analogous
to (\ref{4.15})
\begin{equation}\label{4.21}
\int_{({\cal M}^\epsilon)^{\pm}}\sqrt{-g} d^4x
E_\phi {\cal L}_t\phi=
\pm H^\epsilon_{(m)}
+\int_{({\cal B}^\epsilon)^\pm}
\sqrt{-h}d^3x u_\mu \theta_{(m)}^\mu(t)~~~.
\end{equation}
The last term in r.h.s. of (\ref{4.21}) is the contribution
from the inner boundary, $u^\mu$ is inward pointing normal
of ${\cal B}^\epsilon$. One can now take the limit
$\epsilon\rightarrow 0$
in (\ref{4.21}).
In this limit $H^\epsilon_{(m)}$ coincides with the Hamiltonian
defined on $\Sigma_t$. When $\epsilon$ tends to zero,
$\sqrt{-h}\simeq \epsilon$ on ${\cal B}^\epsilon$ and
(\ref{4.21}) is reduced to (\ref{4.15}). Thus, in the
presence of a Killing horizon Eq. (\ref{4.18}) preserves
its form and ${\cal H}_{(m)}$ does generate canonical transformations
along $t$.

\bigskip

By using relation (\ref{3.23}) it is not difficult to predict
what will be the field equations if one tries to determine
the time evolution
by the energy rather than by the Hamiltonian.
The Poisson bracket of ${\cal E}_{(m)}$ with
$\phi$ contains an extra term which comes
out from the bracket of $\phi$ and $Q_{(m)}$. The Pauli-Jordan function
$\tilde{G}$ vanishes outside the light cone.
For this reason, for the field in the black hole
exterior the extra term is not
zero only on the horizon.
Thus, the time evolution generated by the energy
is different from that generated by the Hamiltonian. The former
corresponds to equations of motion modified by the term
which can be interpreted as a specific
interaction of  fields with the horizon due to
the non-minimal coupling.

\section{${\cal E}_{(m)}$ and black hole thermodynamics}
\setcounter{equation}0

We now show that ${\cal E}_{(m)}$ is the energy of
matter fields which is related to thermodynamical properties
of a black hole.  To this aim we compare a black hole with
zero matter fields (an analog of a vacuum configuration)
to the corresponding black hole with "excited" fields $\phi$.
We will assume that the contribution of the fields is so small
that the back reaction effect can be described by linearized
equations. In what follows we denote with a bar all quantities
for a black hole with $\phi\neq 0$.
Let $M$ and $\bar{M}$ be the mass of a black
hole with $\phi=0$ and $\phi\neq 0$, respectively.
The black hole
metric obeys the equation
\begin{equation}\label{5.1}
E^{\mu\nu}_{(g)}=-\frac 12 T^{\mu\nu}~~~,
\end{equation}
\begin{equation}\label{5.2}
E^{\mu\nu}_{(g)}={1 \over \sqrt{-g}}
{\delta I_{(g)} \over \delta
g_{\mu\nu}}~~~,
\end{equation}
where $I_{(g)}$ is pure gravitational
part (\ref{3.2}) of the action and $T^{\mu\nu}$ is the stress-energy
tensor (\ref{3.6}) of matter fields. Note,
we consider only equations for the metric, but do not require
matter fields to obey the equations of motion.

In the Einstein
gravity
\begin{equation}\label{5.3}
I_{(g)}={1 \over 16\pi G}\int_{\cal M}\sqrt{-g} d^4x R~~~,
\end{equation} where $G$ is the Newton constant
and Eq. (\ref{5.1}) takes a familiar form
\begin{equation}\label{5.4}
R^{\mu\nu}-\frac 12 g^{\mu\nu}R=8\pi G T^{\mu\nu}~~~.
\end{equation}
For a black hole in vacuum, $T^{\mu\nu}=0$
and $S=S_{(g)}$, where $S_{(g)}$ is determined by
Eq. (\ref{2.21a}). When matter fields
are present $T^{\mu\nu}\neq 0$ and the vacuum black hole
metric $g_{\mu\nu}$ changes to $\bar{g}_{\mu\nu}$.
The variation of a matter field from zero to some value $\phi$
has two effects: the black hole mass $M$ changes to $\bar{M}$
and the black hole entropy $S_{(g)}$ changes to
$\bar{S}_{(g)}$.
In the Einstein gravity $S_{(g)}={\cal A}/(4G)$ and
matter fields result in a change
of the area $\cal A$ of the black hole horizon.
The total entropy when $\phi\neq 0$ is
$\bar{S}=\bar{S}_{(g)}+\bar{S}_{(m)}$, where
$\bar{S}_{(m)}$ is the contribution of the matter fields
due to the non-minimal coupling.

Our aim now is to find the relation
between variation of $M$ and $S_{(g)}$ and the energy ${\cal E}_{(m)}$
of the fields. We assume that
$\Sigma_t$ is an infinite hypersurface and fields vanish
at spatial infinity.
First, let us note that according to (\ref{2.13}) and (\ref{2.16})
the black hole mass $\bar{M}$
can be written as
$$
\bar{M}=-\int_{\Sigma_t}\bar{J}^\mu(t)d\Sigma_\mu-\int_{\Sigma}
\sqrt{\sigma} d^2z
\bar{Q}^{\mu\nu}(t) p_\mu l_\nu +\int_{C_\infty} \sqrt{\gamma}d^2y
\bar{B}\bar{N} $$
\begin{equation}\label{5.5}
=-\int_{\Sigma_t}\bar{J}^\mu(t)d\Sigma_\mu+{\bar{\kappa} \over 2\pi}
\bar{S}+ \int_{C_\infty} \sqrt{\gamma}d^2y \bar{B}\bar{N}~~~.
\end{equation}
The total entropy $\bar{S}$ is defined by (\ref{2.18}).
By using Eqs. (\ref{3.13}) and (\ref{3.19}) we get
$$
\bar{M}-\bar{\cal{E}}_{(m)}=
-\int_{\Sigma_t}\left(\bar{J}^\mu(t)-\bar{\theta}_{(m)}^\mu(t)
+t^\mu \bar{L}_{(m)}\right)
d\Sigma_\mu
$$
\begin{equation}\label{5.6}
+{\bar{\kappa} \over
2\pi} \left(\bar{S}+\bar{Q}_{(m)}\right)
+ \int_{C_\infty} \sqrt{\gamma}d^2y \bar{B}\bar{N}~~~.
\end{equation}
We can now use the fact that, according to Eqs. (\ref{3.1})
and (\ref{3.1a})
\begin{equation}\label{5.7}
\bar{J}^\mu(t)=\bar{\theta}_{(g)}^\mu(t)+\bar{\theta}_{(m)}^\mu(t)
-t^\mu(\bar{L}_{(g)}+\bar{L}_{(m)})~~~.
\end{equation}
From (\ref{3.27}) and (\ref{5.7}) it follows that
\begin{equation}\label{5.8}
\bar{M}-\bar{\cal {E}}_{(m)}=
-\int_{\Sigma_t}
\left(\bar{\theta}_{(g)}^\mu(t)-t^\mu \bar{L}_{(g)}\right)
d\Sigma_\mu
+{\bar{\kappa} \over
2\pi} \bar{S}_{(g)}
+ \int_{C_\infty} \sqrt{\gamma}d^2y \bar{B}\bar{N}~~~.
\end{equation}
Let us compare this expression to that when $\phi=0$.
One easily finds that
\begin{equation}\label{5.9}
\delta M-\bar{\cal {E}}_{(m)}=
-\delta \left[\int_{\Sigma_t}
\left(\theta_{(g)}^\mu(t)-t^\mu L_{(g)}\right)
d\Sigma_\mu\right]
+\delta\left({\kappa \over
2\pi} S_{(g)}\right)
+ \delta\left[\int_{C_t} \sqrt{\gamma}d^2y B N\right]~~~.
\end{equation}
Where $\delta$ denotes the difference between the quantities
corresponding to the two black hole solutions.
For instance,
$\delta M=\bar{M}-M$, $\delta S_{(g)}=\bar{S}_{(g)}-S_{(g)}$.

To proceed with (\ref{5.9}) we neglect by all terms which
are of the second order and higher
in the variation of the
black hole metric $\delta g_{\mu\nu}=\bar{g}_{\mu\nu}-g_{\mu\nu}$.
By taking into account Eqs. (\ref{5.1}) and (\ref{5.2})
we find
\begin{equation}\label{5.10}
\delta \left[\int_{\Sigma_t}t^\mu
L_{(g)}d\Sigma_\mu\right]= -\delta \left[\int_{\Sigma_t}\sqrt{-g}
d^3xL_{(g)}\right]
=\int_{\Sigma_t} \sqrt{-g}d^3x\left[\frac 12
T^{\mu\nu}\delta g_{\mu\nu}-
\nabla_\mu \theta_{(g)}^\mu(\delta g)
\right]~~~,
\end{equation}
where we used the fact that $t^\mu d\Sigma_\mu=-\sqrt{-g}d^3x$.
Obviously, the first term in
the r.h.s.  of (5.10)  can be neglected because it is of the second
order in perturbations.

Other terms in Eq. (\ref{5.9}) can be also transformed.
Note that the vacuum metric is static,
${\cal L}_t g_{\mu\nu}=0$, and so
\begin{equation}\label{5.11}
\delta
\theta_{(g)}^\mu(t)\equiv\delta \theta_{(g)}^\mu({\cal L}_t g)=
{\cal L}_t\left(\theta_{(g)}^\mu(\delta g)\right)~~~.
\end{equation}
Thus, by using Eqs. (\ref{5.10}) and (\ref{5.11})
and leaving only the terms linear in the perturbations
one has
$$
-\delta\left[ \int_{\Sigma_t}
\left(\theta_{(g)}^\mu(t)-t^\mu L_{(g)}\right)
d\Sigma_\mu\right]=
-\int_{\Sigma_t}
\left[{\cal L}_t\left(\theta_{(g)}^\mu(\delta g)\right)-
t^\mu\nabla_\nu\theta_{(g)}^\nu(\delta g)\right]
d\Sigma_\mu
$$
$$
=-\int_{\Sigma_t}
\left[t^\nu\nabla_\nu\theta_{(g)}^\mu(\delta g)-
(\nabla_\nu t^\mu) \theta_{(g)}^\nu(\delta g)-
t^\mu\nabla_\nu\theta_{(g)}^\nu(\delta g)\right]
d\Sigma_\mu
$$
\begin{equation}\label{5.12}
=-\int_{\Sigma_t} \nabla_\nu
\left[t^\nu\theta_{(g)}^\mu(\delta g)-
t^\mu \theta_{(g)}^\nu(\delta g)\right]
d\Sigma_\mu=\int_{C_\infty}\sqrt{\gamma}d^2y N u_\nu
\theta_{(g)}^\nu(\delta g)~~~.
\end{equation}
In the last line we took into account that contribution from the
bifurcation surface $\Sigma$, where $t^\mu=0$, is zero. We also
assume that $u^\nu t_\nu=0$ on $C_\infty$.
Matter fields vanish at $C_\infty$. This
guarantees that
$B=B_{(g)}$ in (\ref{5.9}). Beside that,
\begin{equation}\label{5.13}
\int_{\partial {\cal M}}\sqrt{-h} d^3x u_\mu
\theta_{(g)}^\mu(\delta g)=-\delta\left[
\int_{\partial {\cal M}}\sqrt{-h} d^3x
B_{(g)}\right]~~~.
\end{equation}
As a result of (\ref{5.12}) and (\ref{5.13}), Eq.
(\ref{5.9}) is reduced to
\begin{equation}\label{5.14}
\delta M-\bar{\cal E}_{(m)}=\delta\left({\kappa \over
2\pi} S_{(g)}\right)~~~.
\end{equation}
In the given approximation, $\bar{\cal E}_{(m)}={\cal E}_{(m)}$.
From the condition $\delta t^\mu=0$ on $\Sigma$ it
also follows that $\delta \kappa=0$. Thus, our final expression
can be represented as
\begin{equation}\label{5.15}
\delta M=
{\kappa \over 2\pi} \delta S_{(g)}+
{\cal E}_{(m)}~~~.
\end{equation}
This is the direct
generalization of Eq. (\ref{i6}) of black hole thermodynamics in the
Einstein gravity to
arbitrary diffeomorphism invariant theories.
Let us emphasize that variations of
the metric in (\ref{5.15}) should obey the linearized equations
but they are not required to be static.

The quantity
$S_{(g)}$  depends only on geometrical characteristics of the black
hole solution and it can be interpreted as
the {\it proper} black hole entropy.
As distinct from $S_{(g)}$, the entropy $S$ depends
on the non-minimal couplings on the horizon and it is
not a pure black hole characteristic.
Equation (\ref{5.15}) is an analogue
of the first law of black hole thermodynamics (mechanics) \cite{BCH}.
It relates the change of the total mass $M$
to the change of the entropy $S_{(g)}$
and the energy ${\cal E}_{(m)}$ of the matter in the black hole
exterior.

It also follows from (\ref{5.15}) that the energy ${\cal E}_{(m)}$  can be
found by studying the back-reaction effects caused by
excitations of matter fields. To this aim, one has to find
the variation of the black hole mass at spatial infinity and determine
the variation of the geometry near the black hole horizon.
To put it in another way: {\it the energy ${\cal E}_{(m)}$
connects the
change of the black hole mass $M$ with the change of the
proper black hole entropy, i.e. the quantity at spatial infinity
with the quantity at the horizon}.

It should be emphasized that in derivation of (\ref{5.15})
only equations of motion for the metric were used. The matter
fields were assumed to be weak but arbitrary. Thus, Eq. (\ref{5.15})
is a generalization of the first law (\ref{2.19}) to the case
where matter fields are off shell.

\section{Rotating black hole space-times}
\setcounter{equation}0

We now comment on a generalization of
our results to space-times ${\cal M}$ with
axial symmetry and to the case of rotating black holes in particular.
Let $\varphi^\mu$ be a Killing vector
which generates rotations. A conserved charge associated to
the rotational symmetry is the angular momentum of the system.
In the analogy with the Hamiltonian and the energy  it is
possible to give two different definitions of the angular
momentum of matter fields on an external background.
We consider the {\it canonical} angular momentum
${\cal J}^C_{(m)}$
\begin{equation}\label{6.1}
{\cal J}^C_{(m)}=\int_{\Sigma_t} (\theta_{(m)}^\mu(\varphi)-\varphi^\mu
L_{(m)})d\Sigma_\mu~~~,
\end{equation}
and the angular momentum
${\cal J}^E_{(m)}$ defined in terms of the stress-energy tensor
\begin{equation}\label{6.2}
{\cal J}^E_{(m)}=-\int_{\Sigma_t}T^{\mu\nu}\varphi_\mu d\Sigma_\nu~~~.
\end{equation}
Note that the form $\theta_{(m)}(\varphi)$
depends only on the variations ${\cal L}_\varphi \phi$
of the matter fields.
Variations of the background metric vanish ${\cal L}_\varphi
g_{\mu\nu}=0$.
It follows from (\ref{3.11}) that the difference between
two angular momenta is determined by the Noether
current $J_{(m)}^\mu(\varphi)$ associated to the Killing
vector $\varphi^\mu$
\begin{equation}\label{6.3}
{\cal J}^C_{(m)}-
{\cal J}^E_{(m)}=
-\int_{\Sigma_t}
J_{(m)}^\mu(\varphi)d\Sigma_\mu~~~.
\end{equation}
For a space-time without horizons the canonical angular momentum
${\cal J}^C_{(m)}$ is the generator of rotations along
the vector $\varphi^\mu$.  The proof of this fact is analogous
to the proof that
${\cal H}_{(m)}$ generates time translations, see Section 4.

When ${\cal M}$ is a black hole space-time
\footnote{For a discussion of
rotating black holes see Refs. \cite{MTW} and \cite{FrNo:89}.}
additional comments are in order.
The black hole horizon is determined
as a region where the Killing vector $\xi^\mu=t^\mu+\Omega_H
\varphi^\mu$ is null. (Here $\Omega_H$ is the angular velocity
of the horizon).
The bifurcation surface $\Sigma$ is a
region where $\xi^\mu=0$. The black hole horizon is inside of
the static limit surface $S_{st}$
on which the Killing vector $t^\mu$
is null.  $S_{st}$ is the boundary of the ergosphere
where $t^\mu$ is
space-like. Inside the ergosphere ${\cal H}_{(m)}$ and ${\cal E}_{(m)}$
cannot be interpreted as an energy.  For this reason, instead of these
quantities it is more appropriate to consider the conserved charges
corresponding to the vector $\xi^\mu$
\begin{equation}\label{6.4}
{\cal G}^C_{(m)}=-\int_{\Sigma_t}
(\theta_{(m)}^\mu(\xi)-\xi^\mu L_{(m)})d\Sigma_\mu={\cal H}_{(m)}-\Omega_H
{\cal J}_{(m)}^C~~~,
\end{equation}
\begin{equation}\label{6.5}
{\cal G}^E_{(m)}=\int_{\Sigma_t}T^{\mu\nu}\xi_\mu d\Sigma_\nu
={\cal E}_{(m)}-\Omega_H {\cal J}_{(m)}^E~~~.
\end{equation}

The quantity ${\cal G}^C_{(m)}$ is the generator of canonical
transformations along the Killing field $\xi^\mu$ and by repeating
arguments of Section 4 one proves that
for a matter field $\phi$  on the black hole exterior
\begin{equation}\label{6.6}
{\cal L}_\xi \phi(x)=({\cal G}_{(m)}^C,\phi(x))~~~.
\end{equation}
Generalization of relation (\ref{3.23}) to the case of rotating
black holes is
\begin{equation}\label{6.8}
{\cal G}^C_{(m)}-{\cal G}^E_{(m)}=
({\cal H}_{(m)}-{\cal E}_{(m)})-\Omega_H
({\cal J}^C_{(m)}-{\cal J}^E_{(m)})={\kappa \over 2\pi}
Q_{(m)}~~~.
\end{equation}
When the fields vanish on the spatial boundary, the Noether charge
$Q_{(m)}$ is given by Eq. (\ref{3.24}).

\bigskip

Finally, one can find a generalization of the
first law (\ref{5.15}).
By assuming that $C_\infty$
is everywhere tangent to $\varphi^\mu$ it is not difficult to
show that
\begin{equation}\label{6.7}
\delta M={\kappa \over 2\pi} \delta S_{(g)}+\Omega_H\delta{\cal J}
+{\cal E}_{(m)}-\Omega_H{\cal J}^E_{(m)}~~~.
\end{equation}
To prove this one has to
repeat the analysis of Section 5 by replacing ${\cal E}_{(m)}$ by
${\cal G}_{(m)}$ and $t^\mu$ by $\xi^\mu$.
From (\ref{6.7}), we see that the angular momentum
${\cal J}^E_{(m)}$ together with the energy is
related to thermodynamical properties of a black hole.

\section{Summary and discussion}
\setcounter{equation}0

We have shown that the two
definitions of the energy of matter fields in
the presence of a black hole correspond to
different objects. The Hamiltonian ${\cal H}_{(m)}$
is the generator of the time evolution, while the
energy ${\cal E}_{(m)}$ is the quantity which
appears in the first law of black thermodynamics.
The difference between
${\cal H}_{(m)}$ and ${\cal E}_{(m)}$ is the Noether
charge $Q_{(m)}$ which is not zero when fields
are non-minimally coupled.  We derived a formula for
$Q_{(m)}$ valid for an arbitrary diffeomorphism invariant
theory and demonstrated its relation to the Noether charge
introduced by Wald \cite{Wald:93}. As a by product, we found
out the first law of black hole thermodynamics
in case of weak off-shell matter fields. This may be
considered as a further development of
the results obtained in Refs. \cite{Wald:93},\cite{IW:94}.
Equations (\ref{5.15}) and (\ref{6.7}) may be
especially
useful for studying black hole thermodynamics in the
presence of quantum fields, i.e. with the renormalized
quantum stress-energy tensor in the r.h.s. of
(\ref{5.1}).

Now a comment about another possible application of these
results is in order.  As we pointed out,
non-minimally coupled fields are crucial for constructing
ultraviolet finite models of induced gravity.
Calculations of \cite{FFZ},\cite{FF:98}
show that the entropy $S^{BH}$ of a static black hole
in induced gravity is related to statistical-mechanical entropy
$S^{SM}$
and the Noether charge of non-minimally coupled
constituents by Eq. (\ref{i7}).
This can be explained as follows \cite{FF:97}.
The black hole entropy
is connected with the spectrum of the black hole mass $M$.
According to Eq. (\ref{5.15}), if the geometry
near the horizon is fixed the spectrum of $M$ is equivalent
to the spectrum of the energy of matter fields.
On the other hand, $S^{SM}$ is computed by using
the canonical Hamiltonian. Thus,
the entropies $S^{BH}$ and $S^{SM}$ are related to the
different energies and for this reason they do not
coincide.

Our results give a strong support to the above interpretation
of formula (\ref{i7}). The results concern arbitrary
diffeomorphysm invariant theories and suggest that
Eq. (\ref{i7}) is universal and does not depend on the
choice of the concrete induced gravity model. If the
model is ultraviolet finite $S^{BH}$ and $S^{SM}$
always differ by the Noether charge of non-minimally coupled
constituents. Moreover, by taking into account the results
of Section 6, we may speculate that (\ref{i7})
holds for rotating black holes as well. It would be
interesting to check this hypothesis by computations.

\vspace{12pt}
{\bf Acknowledgements}:\ \ I am very grateful to
Valeri Frolov for helpful discussions and
for the hospitality  during my stay in the
University of Alberta, Canada.

\end{document}